\documentclass[12pt,A4paper]{article}
\usepackage{theorem}
\usepackage{makeidx}
\usepackage{amsmath}
\usepackage{amssymb}
\usepackage{epsfig}
\usepackage{graphics}
\usepackage[latin1]{inputenc}
\begin{document}
\noindent
\renewcommand{\thefootnote}{\fnsymbol{footnote}}
\thispagestyle{empty}
\begin{center} {\Large{A Percolation-Based  Model of New-Product Diffusion with Macroscopic Feedback Effects}}

\vspace{1.5cm}

{{Martin Hohnisch, Sabine Pittnauer and Dietrich Stauffer}}{\footnote{Address: Hohnisch: Research Group Hildenbrand, Department of Economics, University of Bonn,
Lenn\'estr. 37, 53113 Bonn, Germany and Research Center BiBoS, University of Bielefeld, D-33501 Bielefeld, Germany 
(e-mail: Martin.Hohnisch@wiwi.uni-bonn.de);
Pittnauer: Research Group Hildenbrand, Department of Economics, University of Bonn,
Lenn\'estr. 37, 53113 Bonn, Germany (e-mail: Sabine.Pittnauer@wiwi.uni-bonn.de);
Stauffer: Institute of Theoretical Physics, University of Cologne, Z\"ulpicher Str. 77, 50923 K\"oln, Germany (e-mail:
stauffer@thp.uni-koeln.de)}}

\end{center}

\vspace{4cm}

{\abstract{
\noindent
This paper proposes a  percolation-based model of new-product diffusion in the spirit of Solomon et al. (2000) and 
Goldenberg et al. (2000).
A consumer buys the new product if she has formed her individual valuation of the product (reservation price) and if 
this valuation is greater or equal than the price of the product announced by the firm in a given period. 
Our model differs from  previous percolation-based models of new-product diffusion in two respects. 
First, we consider macroscopic feedback effects affecting  the supply or the demand side of the market (or both). 
Second, a consumer who did not buy the product in the period in which her valuation was formed remains a potential buyer
and buys in some later period if and when  her individual valuation equals or exceeds the price of the product. 
Unlike most previous 
models of new-product diffusion, our framework accounts for the empirical finding of long  tails   characteristic 
for early stages of innovation diffusion.}}

\vspace{1.3cm}

\noindent
{\it{Keywords:}} Social Percolation, Innovation Diffusion, New-Product Diffusion 
\newpage
\section{Introduction}
\renewcommand{\thefootnote}{\arabic{footnote}}
\setcounter{footnote}{0}

The analysis of the process of new-product diffusion is an important research area in  both marketing science and
economics. From the former perspective, the emphasis is on 
forecasting future sales for a new product depending on the use of elements of the marketing mix
(see e.g. \cite{B2000}). From the latter perspective, that analysis might  
establish some generic features of the dynamics of individual preferences, an issue which has received only little 
attention in economic theory so far.

An aggregate model of new-product diffusion  widely accepted in marketing and business science is due to 
Bass \cite{Ba1969}.  This  model has seen numerous refinements over the years (for an overview, see 
\cite{Ma1990}) and can reproduce 
the evolution of sales over a wide range of the product life cycle, essentially employing a two-parameter fit. 
However, the Bass model and its refinements fail to account for the early stages of the diffusion process in which
sales often fail to raise distinctively  over a prolonged period of time (cf. \cite{Ma1990}, p.21 and \cite{Go1997}).

In recent years, stochastic micromodels of new-product diffusion have begun to emerge (see e.g. \cite{Go2000}). 
In the most simple setting, each consumer is assumed to form independently her valuation (or expectation about the 
product's quality) of the new product and to buy if the price (quality) of the product is smaller (greater) than her 
valuation (expected quality), respectively.

Our approach follows the basic framework of social percolation \cite{So2000}  but
differs from  previous percolation-based models of new-product diffusion in two respects. First, we consider macroscopic 
feedbacks  affecting  the supply or demand side of the market (or both). In the former case, the price of the
product decreases with the cumulative quantity of the product already sold in earlier 
periods reflecting decreasing production costs.  In the latter case, the individual valuation of the product increases 
with the proportion
of other consumers already using the product. Thus, in both cases the probability of an individual to buy the product increases over time. 
Second, a consumer who did not buy the product in the period in which her valuation was formed remains a potential buyer
and buys in some later period if and when  her individual valuation equals or exceeds the price of the product. 

We find that  our framework provides a possible explanation for the empirical finding  of long tails  in early stages
of innovation diffusion and is, up to our knowledge, the first instance of a percolation-based model to do so.

The structure of the paper is as follows. In Section 2 we present the specification of  our model. 
Section 3  provides some generic aspects of the resulting  dynamics. 

\section{The model}

We model the diffusion of a new product (consumer durable) within a large population $A$ consisting of $N$ consumers. 
Each consumer might buy either one unit of the product or none.
Time is discrete and indexed by the natural numbers.

First, the specification of  our  model without macroscopic feedback effects is presented.
If  consumer $a\in A$ has learned in some period $t_a$ about the new product, either from global information sources, e.g. 
the media, or from other consumers who had already bought the product in earlier periods, she forms her  individual
valuation reflected in the reservation 
price $p_a$ (i.e. the highest price at which she would buy). We assume that $p_a$ is a realization of the random variable
$R_a$ with values in $[0,1]$ and probability density function $r_a$. Throughout this paper we assume that $(R_a)_{a\in A}$
is an independently identically distributed family of random variables and $r$ will denote the  probability density 
function common to all consumers.

We use a simple specification of the supply side. We assume that there is a single firm producing the new product and it 
uses mark-up pricing (see e.g. \cite{Ha1988}\cite{Bl1991}\cite{Ha1997}), i.e. the price  $p_t$  of the product in period $t$ equals
\begin{equation}
p_t=(1+m)c_t
\end{equation}
with $c_t$ denoting the unit production costs in period $t$ and $m$ a positive number, the time-constant mark-up.  

To model a consumer's decision to buy the new product, we assume that consumer $a$ buys the product in period 
$t$ with $t\geq t_a$ if 
$p_{a} \geq p_t$ and $p_{a} < p_\tau$ $\forall \tau: t_a\leq \tau < t$.

Furthermore, we need to specify  how information about the existence and the characteristics
 of the product spreads out in the population of consumers, as  a consumer can form her valuation
of the product only if she has obtained this information.{\footnote{In the present model, we assume that the information
passed from one consumer to another is objective in the sense that the valuation formed by the recipient of the information
does not depend on the valuation of the provider of the information.}} 
An important part of that specification is the choice of an appropriate graph structure representing the underlying social 
communication network.  In the present paper  we restrict our attention to the case of a two-dimensional integer-lattice.

We specify the following structure of information flow. 
In the first period, there appears a fixed number of buyers located  randomly over the lattice who are assumed
to have learned about the product from global information sources. In each of the following periods, the neighbours  
of the consumers who have bought the product in the immediately preceeding period  form their valuations.{\footnote{We remark
that some of the neighbours of a new buyer might have already formed their valuations in some earlier period as neighbours
of some other buyer. In that case, the previous valuations remain unchanged.}}
Each consumer who has formed a valuation, either in the present or in previous periods (and has not yet bought the product),
then decides  whether  to buy or not depending on the price announced by the producer. Thus, unlike in the standard
Leath-algorithm  used for the simulation of percolation models, a site once declared inactive may be activated later on
(this aspect will be relevant in the framework with feedback effects).

Up to this point, our specification corresponds to a standard percolation  model. Now we turn to 
macroscopic feedback effects  affecting  supply and demand.
In modelling the feedback affecting the supply side, we assume that unit production costs decrease with the cumulative 
quantity of units already produced. 
The decrease of  unit production costs is empirically well established and explained by learning within the firm.
Decreasing unit production costs are  associated with the notion of the ``learning curve'' in  
business science  \cite{Ye1979}\cite{Ar1990} and with the related notion of ``economies of scale'' in  economics 
(for an overview see e.g. \cite{Sc1990}).
The effect will be quantified by a functional  relationship
$c_t = f( x_{t-1})$
with $x_{t-1}$ denoting the fraction of consumers who  bought the product up to period $t-1$ and $f$ an empirically 
grounded function (which will be specified numerically in Section 3) with $f'(x)<0$ and $f''(x)>0$ for the non-negative real numbers. 
Thus, by Equation (1) we have
 \begin{equation}
p_t=(1+m) f(x_{t-1})
\end{equation}
We model the feedback affecting the demand side  by assuming that  the  reservation price  $p_a$  is augmented
by an amount proportional to  $x_{t-1}$ reflecting the notion of ``network externalities''  increasing the utility 
of a product with the number of other users \cite{Da1985}\cite{Ka1985}\cite{Ka1992}. Thus, in general,  we have a time-dependent reservation price
\begin{equation}
p_{a,t}=p_a + \mu x_{t-1},
\end{equation}
with some constant $\mu$ which we assume to be independent of $a$ for simplicity.

Depending on the nature of the product considered, either one of the feedback effects might vanish. For instance, computer 
software presumably exhibits only  the second kind of feedback effect, while  household electronics  only the first.
Moreover, the reader will notice that both types of macroscopic feedback effects are mathematically equivalent in the 
sense that with increasing  $x_{t-1}$ both narrow the gap between the price of the product
and   the reservation price of consumers who have not yet bought the product. 
For that reason, all our qualitative results remain unchanged if only one of the effects is considered at a time.

\section{Some results on the dynamics of new-product diffusion}
This section presents some  generic features,  obtained by
Monte Carlo simulation, of the evolution of sales resulting from our model with 
the individual reservation price distributed in $[0,1]$.{\footnote{With a 
truncated normal distribution on $[0,1]$, we obtain results which do not differ qualitatively from the ones presented.}}
We restrict ourselves to those features of our model which are
novel as compared with previous percolation-based new-product diffusion models.

Serving as a prototypical example of the long-tailed diffusion curve, Figure 1 (top) depicts the cumulative number of adopters
 of a novel agricultural technique in Iowa. The data in Figure 1 (top) is adapted from \cite{Ry1943}. More examples
of long-tailed diffusion curves can be found in \cite{Go1997}.
Figure 1 (bottom) depicts a diffusion curve resulting from our model with macroscopic feedback affecting the demand side only 
for a setting with one initial buyer 
in period $t=1$, a $400\times 400$ lattice, a time-independent price $p=0.433$ and  the parameter $\mu$ equal  to $0.4$.  
The data is averaged over $500$ simulation runs. 
The reader may think of this averaged curve as modelling new-product diffusion in a population 
located in many towns  with network externalities affecting the population within a single town only.

A comparison of Figure 1 (top) and (bottom) illustrates that our model can  explain  long flat tails empirically observed
in the early stages of new-product diffusion.
 To our knowledge, this particular feature of new-product diffusion has not  been  
obtained from percolation-based micromodels so far.

\begin{figure}[htp]
\begin{center}
\includegraphics [angle=-90,scale=0.5]{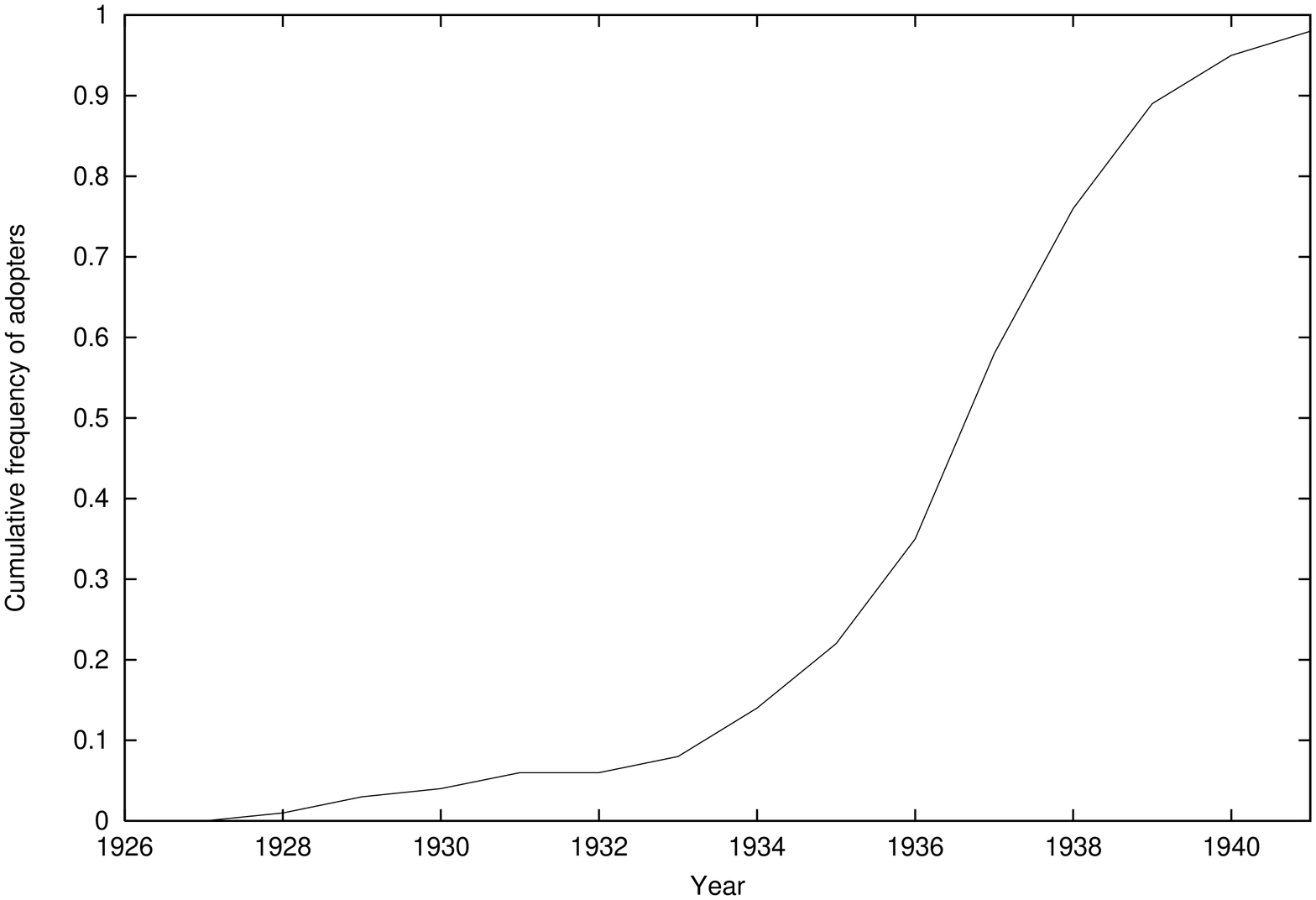}
\includegraphics [angle=-90,scale=0.5]{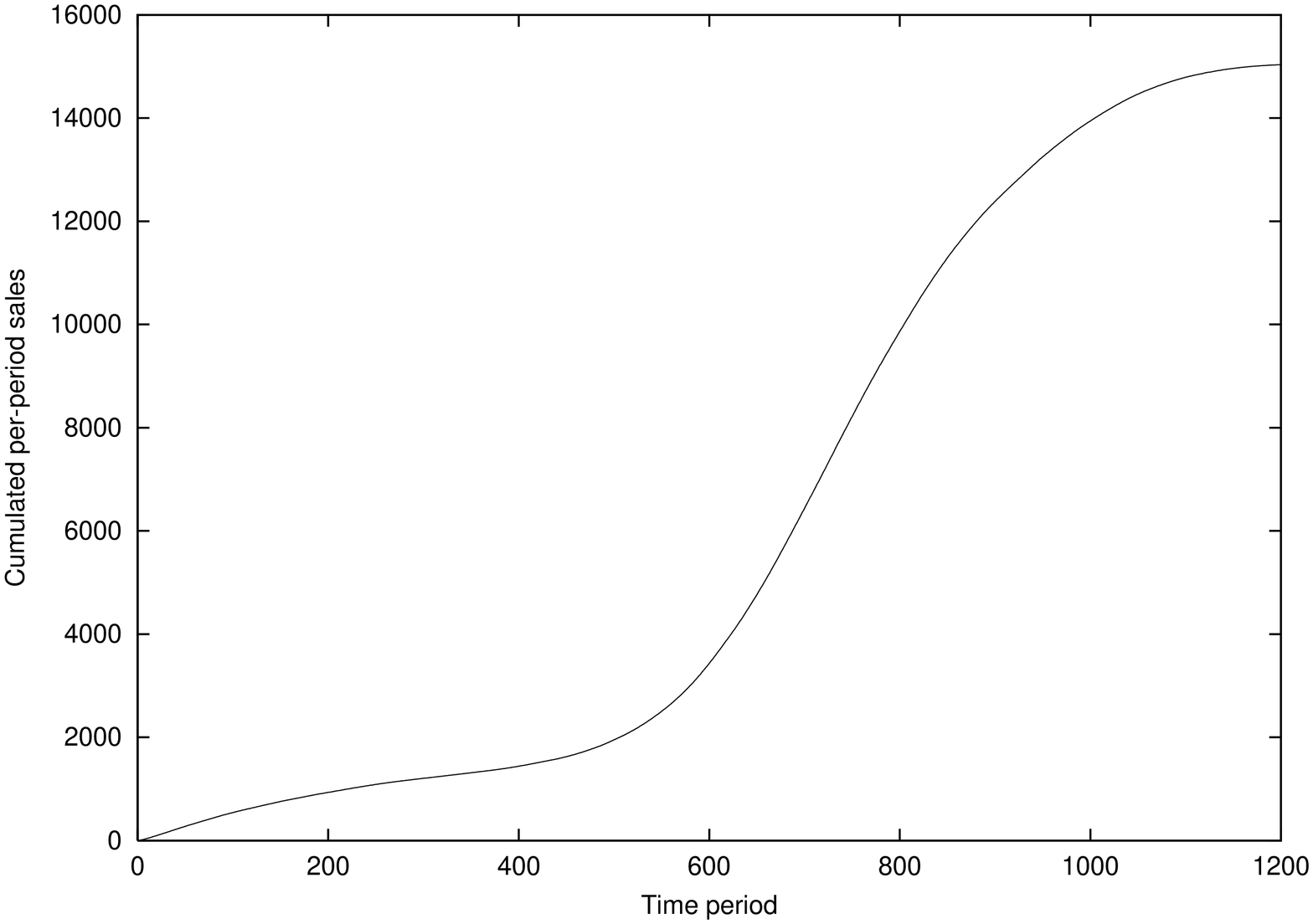}\\
\caption{\label{kumsum} Cumulative frequency of adopters for the diffusion of hybrid corn seed in 
two Iowa farming communities adapted from \cite{Ry1943} (top); cumulative
number of adopters resulting from our model (bottom) with the parameter values 
$p=0.433$ and  $\mu=0.4$.}
\end{center}
\end{figure}

The feature  of long tails in the early stages of the new-product diffusion, visible  in Figure 1 (bottom), 
obtains in our model  for a wide range of parameters and
initial settings. Figures 2 and 3 provide a more detailed characterization  of that feature.

Figure 2 depicts  the evolution of  per-period sales (left-hand side) and cumulative sales (right-hand side) resulting
 from   a specification with one initial buyer in period $t=1$ on a $400 \times 400$ lattice.  
Macroscopic feedback affect the demand side only; the time-independent price is set to 
 $p=0.435$ (top) and  $p=0.421$ (bottom) and  the constant $\mu$ describing  the influence of network 
externalities equals  $0.4$. For both prices, the curves are obtained by averaging over $500$ simulation runs. 

Note that the product price corresponding to the site-percolation threshold for the underlying model on a two-dimensional
 lattice equals  approximately $0.407$ as a consumer buys if her reservation price, uniformly  distributed on $[0, 1]$,
 equals or exceeds the product price.
Thus, for both  prices  in Figure 2 percolation  would not occur in the standard percolation model.
However,  because individual reservation prices increase with the proportion of buyers  $x_{t}$ 
 some simulation runs persist up to the point where percolation occurs. 
 Thus, averaged over $500$    runs, the per-period sales curve exhibits two specific phases.
First, a slow take-off, in some cases even displaying  a temporary decrease
of per-period sales, resulting from runs who terminate before reaching the percolation threshold.
Second, a distinct rise  following the slow start resulting  from those runs which reached the percolation threshold.

The length of the long left tail increases  with $p$ and decreases with $\mu$ ceteris paribus, respectively. 
A comparison of Figure 2 (top) with Figure 2 (bottom) exemplifies the first part of this statement. Furthermore, 
the decrease of per period sales in the first phase as visible in Figure 2 (top, left-hand side), is amplified with increasing
price, ceteris paribus.

\begin{figure}[htp]
\begin{center}
\includegraphics[angle=-90,scale=0.29]{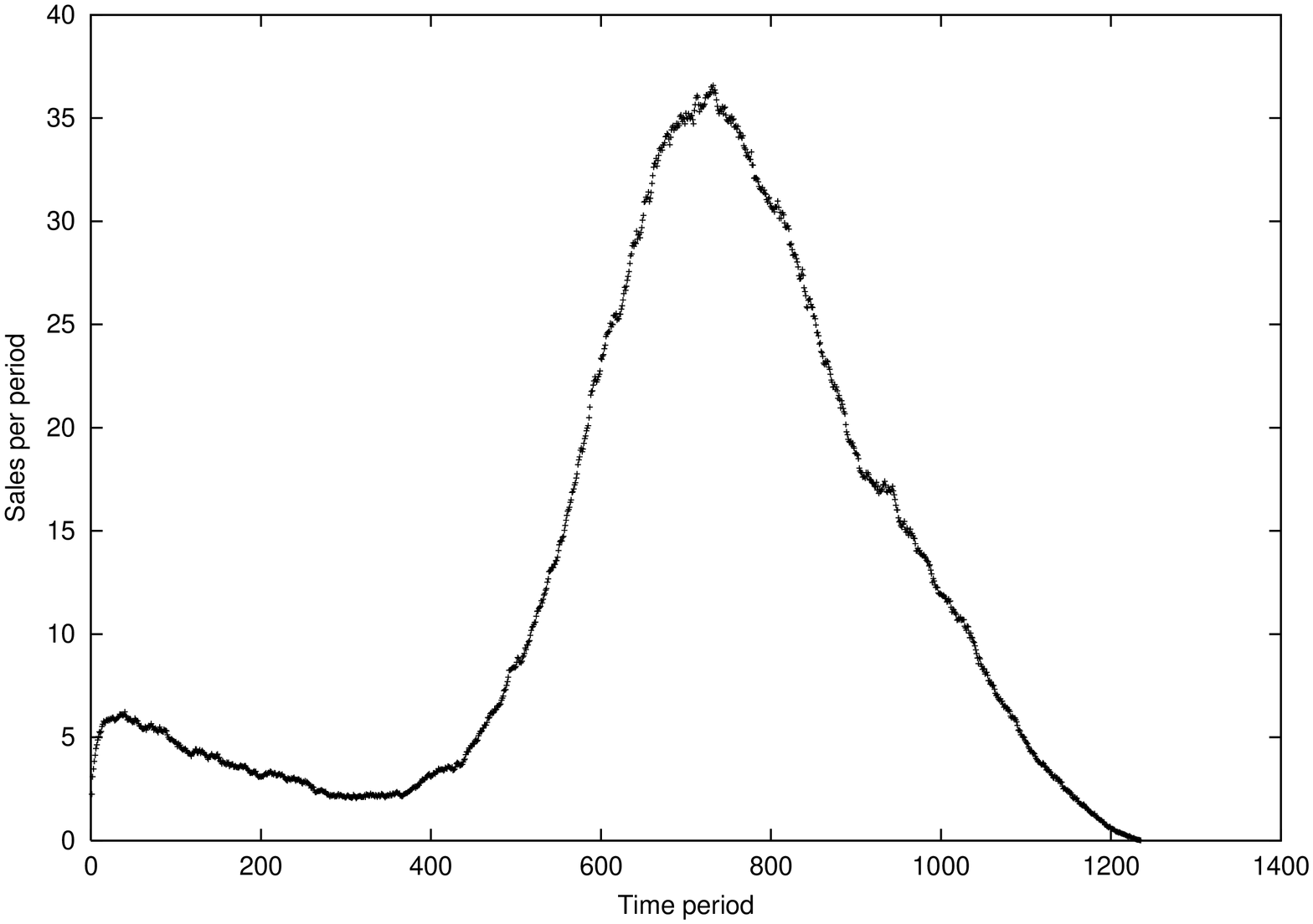}
\includegraphics [angle=-90,scale=0.29]{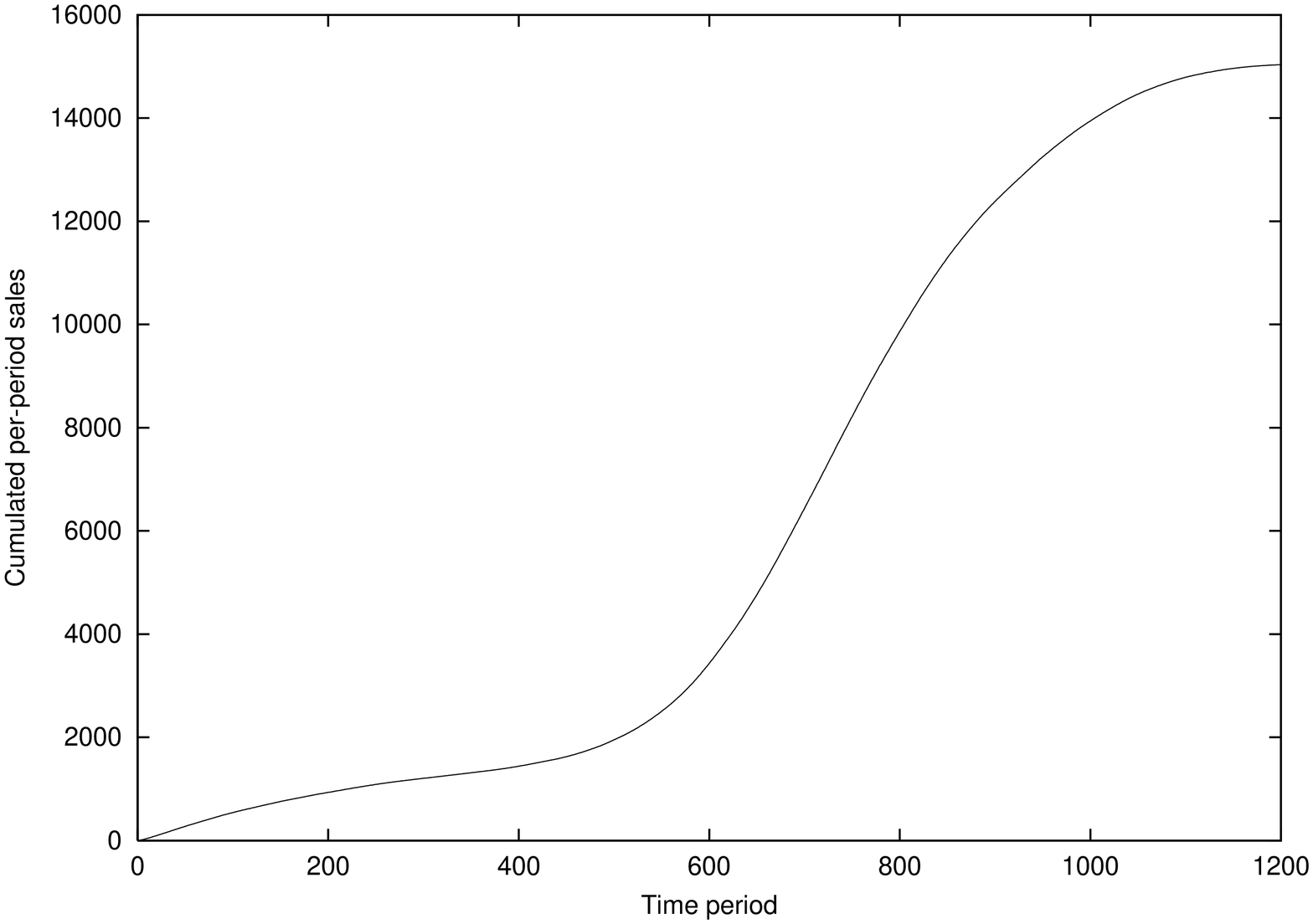}
\includegraphics [angle=-90,scale=0.29]{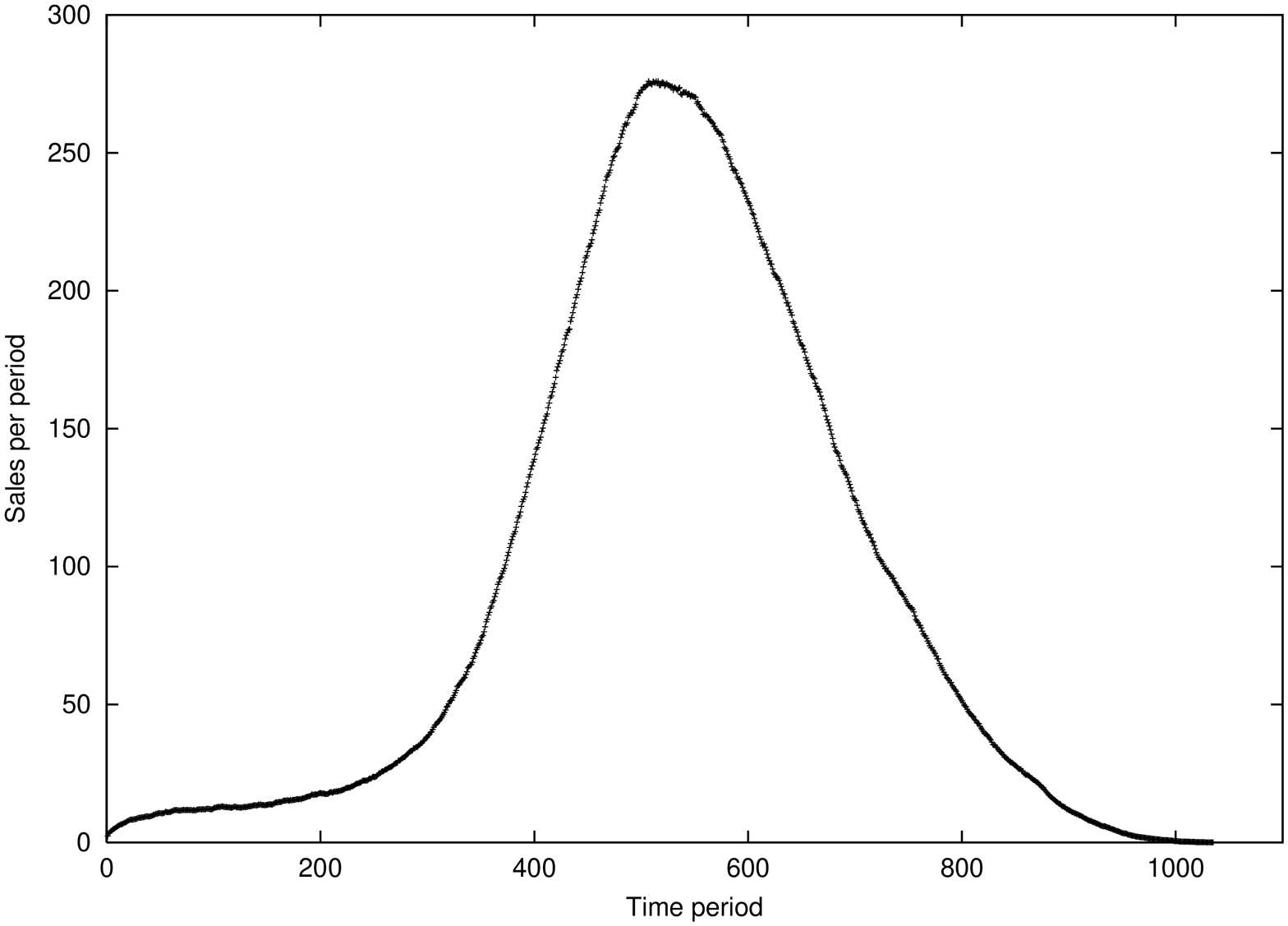}
\includegraphics [angle=-90,scale=0.29]{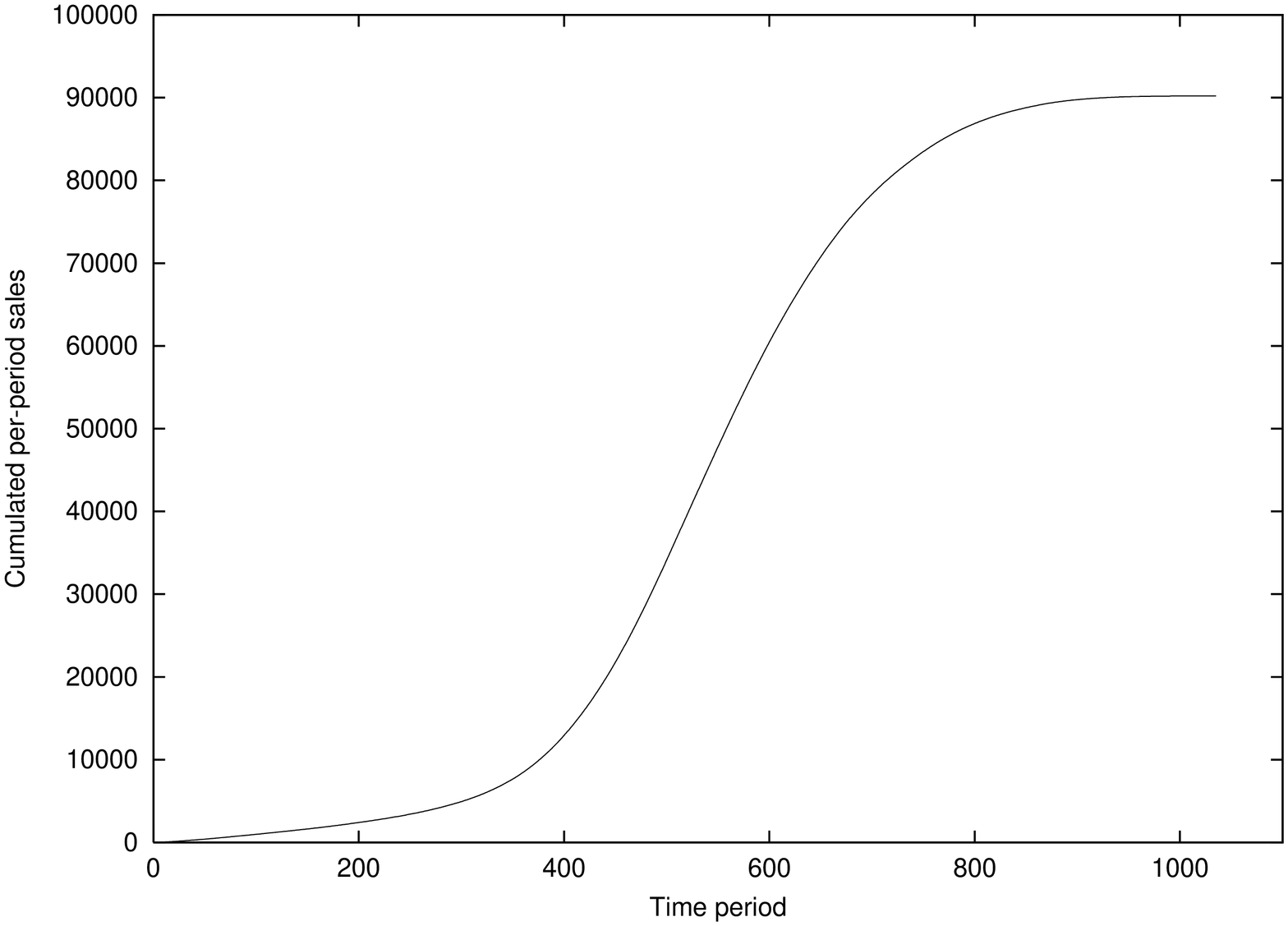}\\
\caption{\label{frequency} Evolution of  per-period sales (left-hand side) and  cumulative sales (right-hand side) 
with  parameter values  $p=0.435$ and $\mu=0.4$ (top) and  $p=0.421$ and $\mu=0.4$ (bottom).}
\end{center}
\end{figure}

\vspace{-0.5cm}

Figure 3 shows three curves corresponding to per-period sales, cumulative sales and the evolution of the product price
resulting from  a single simulation run in a setting with macroscopic feedback affecting the supply side only.
The initial number of buyers  equals 3000 and lattice size is $1500 \times 1500$. Initial price $p_0$ is set to $0.52$.

We use the following specification of the empirically grounded functional relationship $f$ (see Equation (2)) quantifying
the cost decrease, and thus the decrease of the product price,  with the number 
of product units sold:
\begin{equation}
(1+m)f(x)=p_0 -qx + \alpha x^2.
\end{equation}
In this specification in terms of the fraction of buyers $x_t$ as variable, $p_0$ denotes the initial price, i.e $(1+m)f(0)=p_0$,  
$q>0$ the absolute value of the initial slope of $(1+m)f$, i.e. $(1+m)f'(0)=-q$  and $\alpha$ a constant 
determining the convexity of the price decrease function.  
Figure 3 corresponds to the parameter values $q=0.5$, and  $p_0=0.52$ and $\alpha=0.295$.

As Figure 3 indicates, the  same characteristics of diffusion curves as displayed by the averaged curves of Figure 2 can be
obtained from a single simulation run. This fact is important because a market only allows 
for  one price for a single product and thus averaged curves are difficult to interpret 
 in the case of macroscopic feedbacks affecting price.


\begin{figure}[htp]
\begin{center}
\includegraphics [angle=-90,scale=0.38]{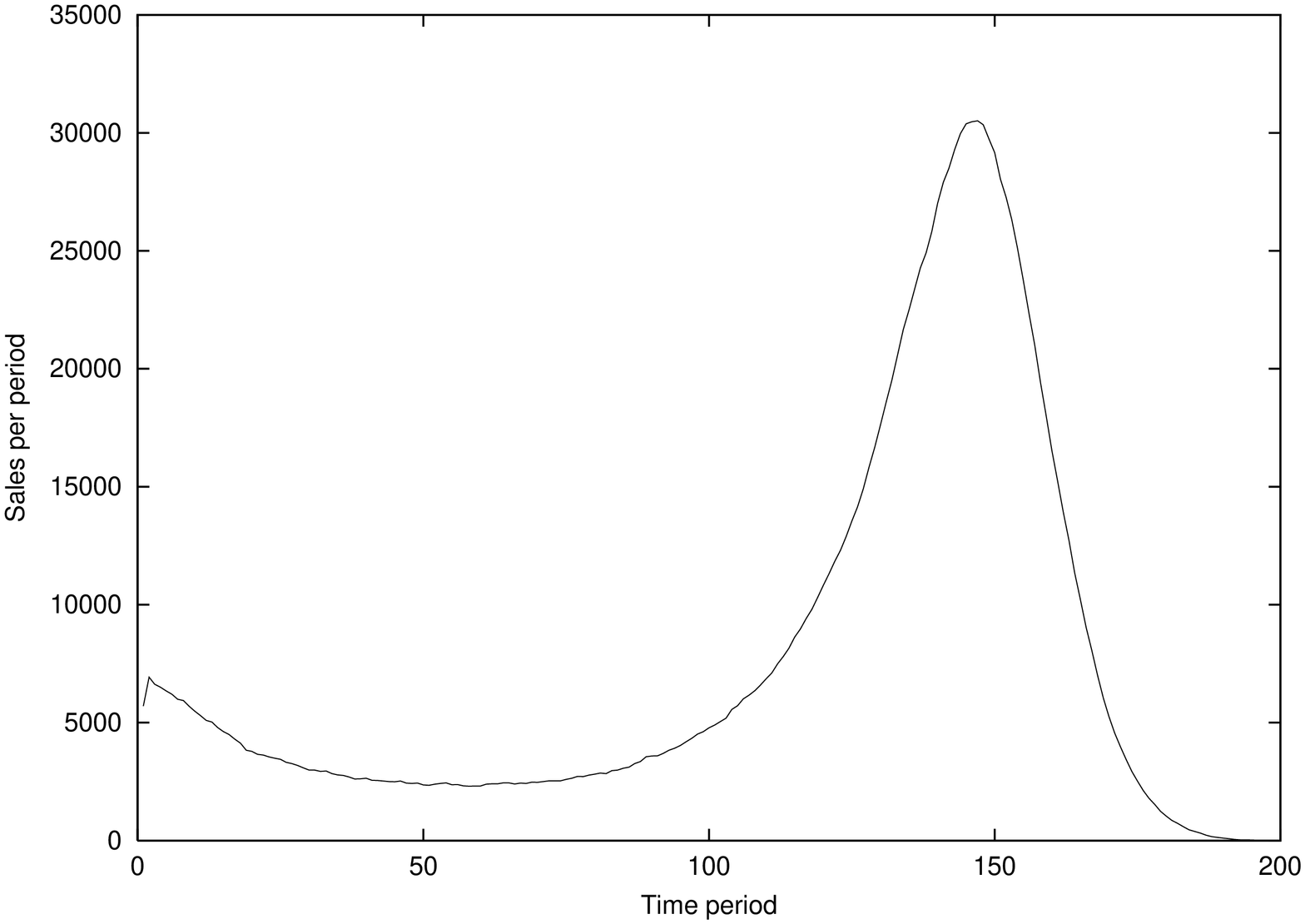}\\
\includegraphics[angle=-90,scale=0.38]{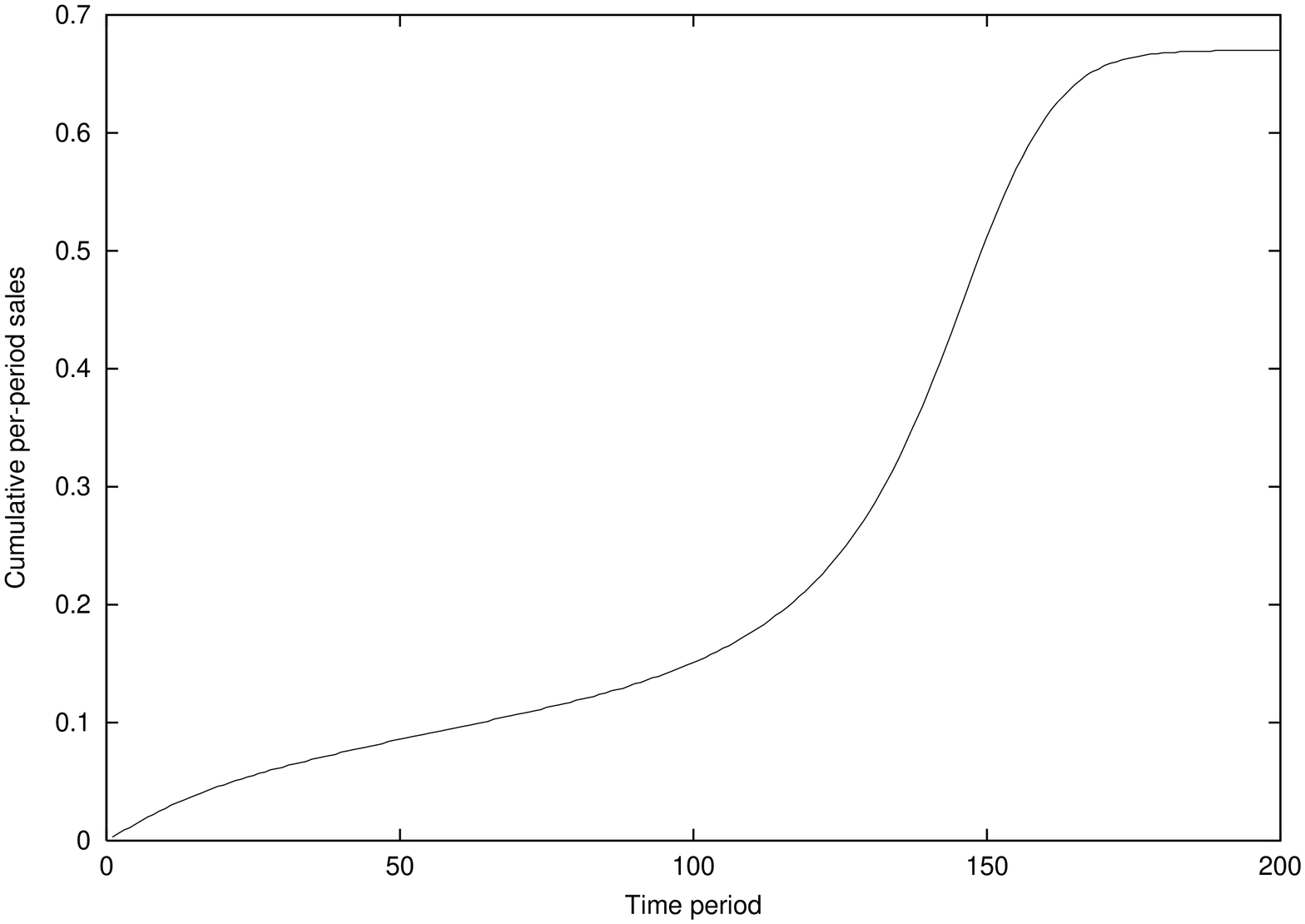}\\
\includegraphics [angle=-90,scale=0.38]{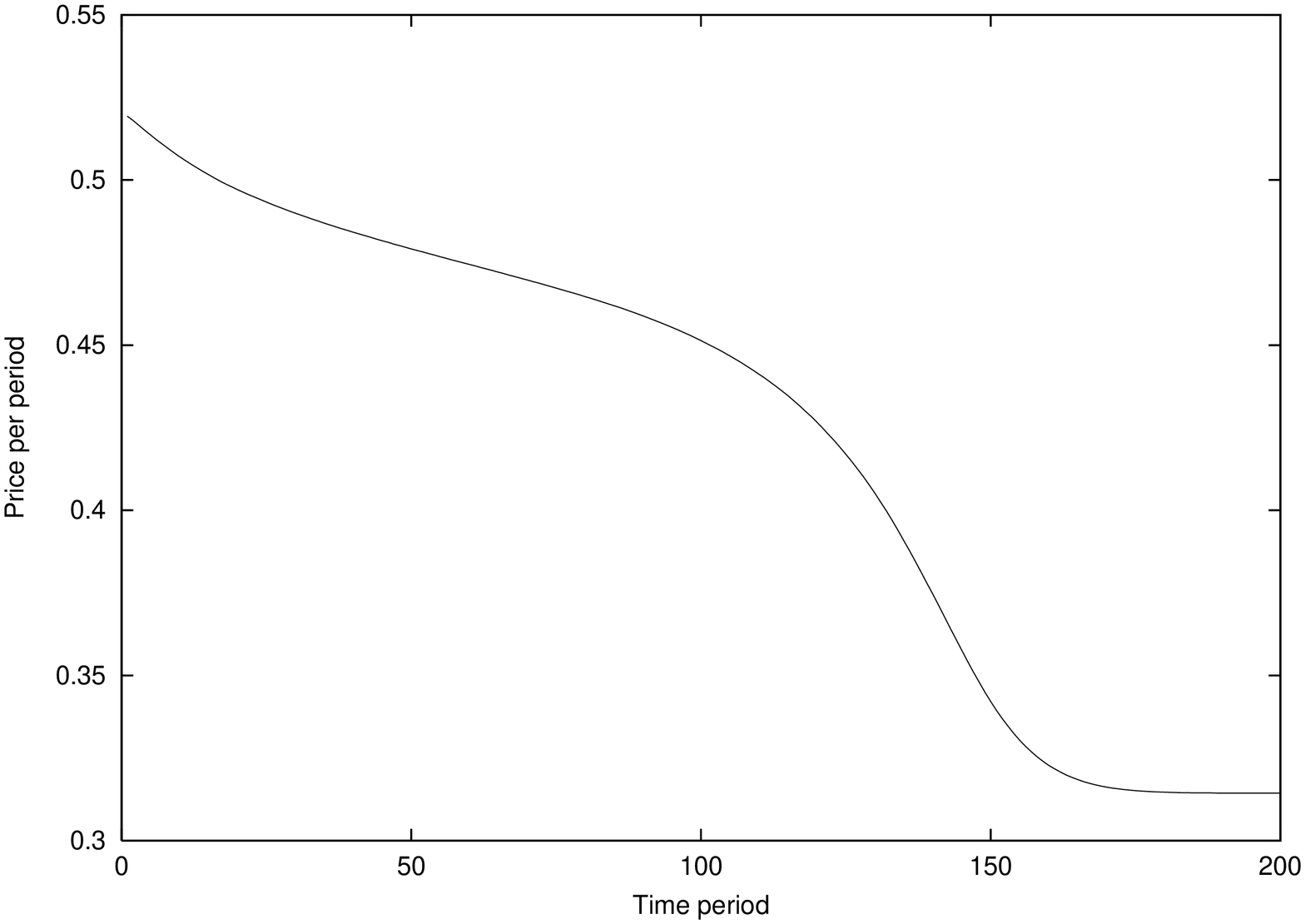}\\
\caption{\label{frequency} Evolution of  per-period sales, cumulative sales and price resulting in a framework with
macroscopic feedback affecting supply side only;
initial price  equals $p_0=0.52$}
\end{center}
\end{figure}


\newpage

\begin{thebibliography}{99}
\bibitem{Ar1990} 
Argote, L., Beckman, S.L. and Epple, D. (1990) The Persistence and Transfer of Learning in Industrial Settings,
{\it{ Management Science}}, Vol. 36 No. 2, pp. 140-154. 
\bibitem{Ba1969} Bass, F.M. (1969), A New Product Growth Model for Consumer Durables, {\it{Management Science}} 15, 215-227
\bibitem{B2000} Bass, F.M. , D. Jain and T. Krishnan (2000), Modelling the Marketing-Mix Influence in New Product Diffusion,
in: {\it{New-Product Diffusion Models}}, ed. V. Mahajan, E. Muller and Y. Wind, Kluver AP
\bibitem{Bl1991} Blinder, A. S. (1991), Why are Prices Sticky? Preliminary Results from an Interview Study,
 {\it{American Economic Review}} 81, 89-96 
\bibitem{Da1985} David, P.A. (1985), Clio and the Economics of QWERTY, {\it{American Economic Review}} 75, 332-337
\bibitem{Go2000} Goldenberg, J, B. Libai, S. Solomon, N. Jan and D. Stauffer (2000), 
Marketing percolation, {\it{Physica A}} 284 (1-4)  pp. 335-347 
 \bibitem{Go1997} Golder, P.N. and G.J. Tellis (1997), Will It Ever Fly? Modelling the Takeoff of Really New
Consumer Durables, {\it{Marketing Science}} 16 (3), 253-270
\bibitem{Ha1988} Hall, R. (1988), The Relation between Price and Marginal Cost in U.S. Industry,
{\it{Journal of Political Economy}}, 96(51), 921-947
\bibitem{Ha1997} Hall, S., Walsh, M. and Yates, A. (1997), How do UK Companies set Prices ?, 
{\it{Bank of England Working Papers}} 67
\bibitem{Ka1985}Katz M.L. and C. Shapiro (1985), Network Externalities, Competition, and Compatibility,
 {\it{American Economic Review}}  75, 424-440
\bibitem{Ka1992}Katz M.L. and C. Shapiro (1992), Product Introduction with Network Externalities, 
 {\it{Journal of Industrial Economics}}  40, 55-84
\bibitem{Ma1990} Mahajan, V., Muller, E. and Bass, F.M. (1990), New Product Diffusion Models in Marketing: A
Review and Directions for Research, {\it{Journal of Marketing}}  54, 1-26
\bibitem{Ry1943} Ryan, B. and Gross N.C. (1943), The Diffusion of Hybrid Seed Corn In Two Iowa Communities, 
{\it{Rural Sociology}} 8, 15-24
\bibitem{Sc1990} Scherer, F.M. and Ross, D. (1990), {\it{Industrial Market Structure and Economic Performance}},
$3^{rd}$ edition, Houghton Mifflin
\bibitem{So2000}Solomon, S., G. Weisbuch, L. de Arcangelis, N. Jan and D. Stauffer 
Social percolation models, {\it{Physica A}} 277 (1-2) (2000) pp. 239-247 
\bibitem{Ye1979}Yelle, L.E. (1979), The Learning Curve: Historical Review and Comprehensive Survey, 
{it{Decision Sciences}} Vol. 10, pp. 302-328. 

\end{thebibliography}
\end{document}